\documentclass[preprint,5p,twocolumn]{elsarticle}
\usepackage{lipsum}
\usepackage{tikz} \usetikzlibrary{positioning} \tikzset{font=\footnotesize}
\usepackage{standalone}
\usepackage{graphicx} 

\usepackage{color}
\usepackage{subcaption} 
\usepackage{makecell} 
\usepackage{amssymb, pifont}

\usepackage{amsmath,amssymb,amsfonts}
\usepackage[ruled, lined, linesnumbered, commentsnumbered, longend]{algorithm2e}
\usepackage{setspace}
\usepackage{amssymb}
\usepackage{tablefootnote}

\usepackage{graphicx}
\usepackage{textcomp}
\usepackage{xcolor}
\usepackage{soul}
\usepackage{enumerate}
\usepackage{comment}

\usepackage{multirow}
\usepackage{hyperref}
\usepackage{tabu}
\usepackage{float}
\usepackage{booktabs}
\usepackage{arydshln}
\usepackage[inline]{enumitem}

\usepackage{amsmath}
\usepackage{mathtools}
\usepackage{textcomp}
\usepackage{xurl}

\usepackage{threeparttable}

\definecolor{boristext}{rgb}{0.22, 0.44, 0.33}
\definecolor{boriscomments}{rgb}{0.83, 0.0, 0.0}
\definecolor{miguelcomments}{rgb}{0.5, 0, 0.8}
\definecolor{migueltext}{rgb}{0.42, 0.1, 0.9}

\begin{document}

\begin{frontmatter}

\title{Performance Evaluation of Multi-Armed Bandit Algorithms for Wi-Fi Channel Access}

\journal{npj Wireless Technology}

\author[upf]{Miguel Casasnovas}
\author[upf]{Francesc Wilhelmi}
\author[supelec]{Richard Combes}
\author[agh]{Maksymilian Wojnar}
\author[agh]{Katarzyna Kosek-Szott}
\author[agh]{Szymon Szott}
\author[upf]{Anders Jonsson}
\author[uab]{Luis Esteve}
\author[upf]{Boris Bellalta}

\address[upf]{Universitat Pompeu Fabra, Barcelona, Spain}
\address[supelec]{CentraleSupélec, Paris, France.}
\address[agh]{AGH University of Krakow, Kraków, Poland.}
\address[uab]{Universitat Autònoma de Barcelona, Barcelona, Spain. \vspace{-2ex}}

\begin{abstract}
The adoption of dynamic, self-learning solutions for real-time wireless network optimization has recently gained significant attention due to the limited adaptability of existing protocols. This paper investigates multi-armed bandit~(MAB) strategies as a data-driven approach for decentralized, online channel access optimization in Wi-Fi, targeting dynamic channel access settings: primary channel, channel width, and contention window~(CW) adjustment. Key design aspects are examined, including the adoption of joint versus factorial action spaces, the inclusion of contextual information, and the nature of the action-selection strategy (optimism-driven, unimodal, or randomized).
State-of-the-art algorithms and a proposed lightweight contextual approach, E-RLB, are evaluated through simulations. Results show that contextual and optimism-driven strategies consistently achieve the highest performance and fastest adaptation under recurrent conditions. Unimodal structures require careful graph construction to ensure that the unimodality assumption holds.
Randomized exploration, adopted in the proposed E-RLB, can induce disruptive parameter reallocations, especially in multi-player settings. 
Decomposing the action space across several specialized agents accelerates 
convergence but increases sensitivity to randomized exploration and demands coordination under shared rewards to avoid correlated learning. 
Finally, despite its inherent inefficiencies from $\epsilon$-greedy exploration, E-RLB demonstrates effective adaptation and learning, highlighting its potential as a viable low-complexity solution for realistic dynamic deployments.
\end{abstract}


\end{frontmatter}


\section{Introduction}

Wi-Fi (IEEE~802.11) networks, now evolving toward Wi-Fi~8, have become the dominant wireless access technology worldwide~\cite{CiscoAIR2020}. However, their medium access control~(MAC) remains largely static and heuristic-driven, relying on preconfigured protocols and parameters that fail to fully adapt to dynamic traffic and environmental conditions (e.g., a variable number of contenders). This rigidity often results in inefficient spectrum utilization, persistent collisions, and significant performance degradation, particularly in dense deployments.

Reinforcement learning~(RL), and more specifically \emph{multi-armed bandits}~(MABs), are seeing increasing adoption as a means to enable adaptive wireless network optimization~\cite{szott2022wi}. MABs provide a lightweight and sample-efficient online learning framework in which an agent repeatedly selects actions and updates its decisions based on observed outcomes or rewards, without requiring \emph{a priori} knowledge of the environment or explicit state modeling. Consequently, MAB formulations have been successfully applied to a variety of wireless network optimization problems---significantly outperforming legacy IEEE~802.11 operation---including channel selection~\cite{maghsudi2014joint, wilhelmi2019collaborative, barrachina2021multi}, transmit power control~\cite{maghsudi2014joint, wilhelmi2019collaborative}, spatial reuse~\cite{wojnar2025ieee, wojnar2025coordinated, wilhelmi2024coordinated}, and online rate adaptation~\cite{8409322, le2025wifi}.

Building upon the MAB framework, \emph{contextual multi-armed bandits}~(CMABs) incorporate contextual or environmental information (e.g., channel occupancy, or traffic load) into the decision process. By conditioning action selection to the observed context, CMABs enable more informed and responsive adaptation while retaining the lightweight, stateless nature of conventional MABs. As such, CMABs have also been proven effective for various wireless optimization tasks, including network selection~\cite{martinez2023contextual} and spatial reuse~\cite{iturria2024cooperate}, where they outperformed non-contextual approaches. 

Building on our previous work~\cite{casasnovas2025paper1}, this paper presents a comprehensive, simulation-based study of MAB and CMAB algorithms for online, decentralized Wi-Fi channel access optimization. The study focuses on the adaptive tuning of key MAC parameters: the channel width, the primary channel, and the contention window~(CW) size. Three representative approaches to action selection are compared---optimism-driven (confidence-based), unimodal (structure-exploiting), and randomized (stochastic exploration)---and evaluated under both joint (single-agent, SA) and factorial (multi-agent, MA) action-space formulations, and in both single- and multi-player settings. The analysis includes state-of-the-art algorithms---Upper Confidence Bound~(UCB)~\cite{auer2002finite}, Linear UCB~(LinUCB)~\cite{li2010contextual}, and Optimal Sampling for Unimodal Bandits~(OSUB)~\cite{combes2014unimodal}---as well as a newly proposed lightweight contextual algorithm, the \emph{Epsilon-RMSProp Linear Bandit}~(E-RLB), which combines randomized exploration ($\epsilon$-greedy) with RMSProp-based online updates~\cite{hinton2012lec6} to learn efficiently in non-stationary environments.
To the best of our knowledge, this is the first comprehensive evaluation of these algorithms across both joint and factorial action spaces for decentralized Wi-Fi channel access. 

The main contributions of this paper are as follows:
\begin{enumerate}
    \item \textit{Proposal of a new algorithm:} The introduction of E-RLB, a lightweight contextual linear bandit algorithm that can serve as a foundation for future CMAB-based designs.
    \item \textit{Comprehensive comparative study:} An extensive evaluation of MAB and CMAB strategies for Wi-Fi channel access across different interaction settings, analyzing their performance, behavior, and limitations.
    \item \textit{Design insights:} Practical insights into how action-space decomposition, context incorporation, and exploration strategies influence learning efficiency and MAC-layer optimization.
\end{enumerate}

The remainder of this paper is organized as follows: Section~\ref{sec:related_work} reviews related work; Section~\ref{sec:background} describes the MAB learning framework, including the action space, context space, and reward design; Section~\ref{sec:algorithms} details the algorithms under study, including E-RLB; Section~\ref{sec:evaluation} presents the considered single- and multi-player scenarios and their results; Section~\ref{sec:takeaways} summarizes the main takeaways; 
and Section~\ref{sec:conclusions} provides final remarks and directions for future work.



\section{Related Work}\label{sec:related_work}

MABs have recently gained significant attention for online wireless network optimization due to their low computational complexity and fast convergence properties. For instance, Maghsudi et al.~\cite{maghsudi2014joint} investigated multi-player MABs for joint channel selection and transmission power control in adversarial settings, demonstrating that self-interested, decentralized agents can reach stable equilibria and exhibit cooperative behavior without explicit coordination. In addition, Wilhelmi et al.~\cite{wilhelmi2019collaborative} studied multiple exploration strategies (e.g., $\epsilon$-greedy, EXP3, UCB, and Thompson sampling) for both channel selection and power control in uncoordinated wireless deployments. The authors showed that even in adversarial multi-player settings, proportional fairness and collaborative behaviors can be achieved. However, they also observed high individual performance variability due to neighboring networks' behavioral dynamics, particularly under randomized exploration strategies such as $\epsilon$-greedy and EXP3.

More recently, Wojnar et al.~\cite{wojnar2025ieee, wojnar2025coordinated} proposed hierarchical MAB models for coordinated spatial reuse and evaluated multiple algorithms ($\epsilon$-greedy, UCB, Thompson sampling, and Softmax). The authors found that UCB achieves a good trade-off between fast convergence and sustained performance, that $\epsilon$-greedy converges quickly but often fails to identify optimal configurations, that Thompson sampling requires longer operation to refine its uncertainty estimates, and that Softmax exhibits notable parameter sensitivity.
Similarly, Wilhelmi et al.~\cite{wilhelmi2024coordinated} designed a coordinated bandit scheme leveraging IEEE~802.11 multi-AP coordination with joint reward structures (e.g., average, min--max) across neighboring basic service sets (BSSs) to achieve spatial reuse. The authors found that explicit exploration ($\epsilon$-greedy) offers controllable exploration rates but may fail to adapt to distinct conditions, whereas implicit exploration (Thompson sampling) provides better adaptability at the cost of transient instability. Barrachina-Muñoz et al.~\cite{barrachina2021multi} evaluated multiple action-selection strategies (exploration-first, $\epsilon$-greedy, UCB, EXP3, and Thompson sampling) for primary-channel and channel-width selection in dense multi-player scenarios. The authors found that lightweight strategies (exploration-first, $\epsilon$-greedy) outperform the other methods under uncoordinated operation, as excessive exploration leads to unnecessary instability. 
In addition, Szczech et al.~\cite{szczech2025toward} demonstrated the practical feasibility of MAB-based MAC-layer adaptation, using Thompson sampling to dynamically adjust CW, request-to-send/clear-to-send (RTS/CTS), and frame aggregation, achieving substantial performance gains over static IEEE~802.11 configurations. Finally, Combes et al.~\cite{8409322} and Le et al.~\cite{le2025wifi} used unimodal bandit approaches for IEEE 802.11 rate adaptation, showing improved performance in simulations and testbed experiments. In particular, Combes et al.~\cite{8409322} formulated online rate adaptation as a graphically unimodal structured MAB problem, outperforming conventional approaches in both stationary and non-stationary environments, whereas Le et al.~\cite{le2025wifi} combined cascaded and unimodal bandit models with Thompson sampling, achieving higher throughput and faster convergence than other Thompson sampling-based methods and Linux's default rate adaptation algorithm.

Beyond non-contextual methods, several works have demonstrated the benefits of incorporating contextual information into the agents operation. 
For instance, Martínez et al.~\cite{martinez2023contextual} applied contextual bandits for network selection, showing that polynomial extensions of LinUCB outperform UCB under both stationary and non-stationary conditions. 
Iturria et al.~\cite{iturria2024cooperate} compared contextual (SAU-sampling and Reward-cooperative $\epsilon$-greedy) and non-contextual ($\epsilon$-greedy, UCB, and Thompson sampling) bandits for power control and clear-channel assessment threshold configuration in cooperative and non-cooperative multi-player settings. Their results demonstrated that sharing the reward among players enhances fairness, that $\epsilon$-greedy yields gains under reduced action sets, and that cooperative contextual methods are highly effective in dynamic scenarios.

Both \cite{martinez2023contextual} and \cite{iturria2024cooperate} highlight the advantages of contextual learning, motivating the use of contextual formulations in this paper and the design of the proposed E-RLB algorithm. Unlike Iturria et al.'s proposal (Reward-cooperative $\epsilon$-greedy), E-RLB operates in a fully decentralized manner, requiring no access to peer information, and employs stochastic gradient descent (SGD) updates, which has proven effective for enabling efficient parameter learning in high-dimensional contexts. For instance, Ding et al.~\cite{ding2021efficient} developed a linear contextual bandit combining Thompson sampling with single-step SGD updates, showing strong theoretical guarantees and empirical efficiency. 
Chen et al.~\cite{chen2021statistical} proposed a non-linear contextual bandit using $\epsilon$-greedy exploration and inverse probability weighting, also based on SGD updates. In contrast, E-RLB builds upon linear contextual bandits (unlike~\cite{chen2021statistical}) with $\epsilon$-greedy exploration (unlike~\cite{ding2021efficient}) and introduces exponential moving average (EMA) smoothing to handle non-stationary environments. This enables a fair comparison of exploration paradigms under a shared contextual framework, i.e., LinUCB (contextual linear bandit using optimism-based exploration) against E-RLB (contextual linear bandit using randomized exploration).

Overall, existing literature has demonstrated the applicability of MABs for wireless network optimization, including MAC-layer parameter adaptation. This motivates our endeavor to optimize Wi-Fi channel access by jointly tuning multiple MAC parameters (primary channel, channel width, and CW) through data-driven learning in both single- and multi-player settings. In contrast to most prior works, which mostly rely on a single agent (either centralized or per-BSS) optimizing all the parameters, we investigate the impact of action-space decomposition, introducing multiple specialized agents per BSS that independently optimize individual parameters. The literature has also established that the exploration strategy critically influences performance and variability in wireless systems, motivating our comparative analysis of distinct exploration paradigms. Moreover, structured bandit formulations such as those in~\cite{8409322, le2025wifi} motivate our investigation of unimodal approaches (e.g., OSUB).

In summary, this paper contributes to the existing literature by (i) introducing E-RLB, a contextual linear bandit for decentralized Wi-Fi channel-access optimization; (ii) explicitly contrasting three exploration paradigms (optimism-driven, unimodal, and randomized); (iii) comparing single-agent (joint) and multi-agent (factorized) learning architectures; and (iv) evaluating both contextual and non-contextual formulations in single- and multi-player environments. 


\section{Learning-Driven IEEE 802.11 Channel Access}\label{sec:background}

\subsection{IEEE 802.11 Operation}
IEEE~802.11 supports aggregating contiguous 20~MHz channels into wider 40-, 80-, 160-, or 320-MHz bands through channel bonding, where one channel acts as \emph{primary} (where contention is done) and the rest as \emph{secondary} (which availability is checked right before initiating a transmission). More specifically, channel access follows the Distributed Coordination Function~(DCF).
Before transmission, the primary channel must remain idle for a fixed period (DIFS), after which a random backoff counter (drawn from $[0, \text{CW}-1]$) starts. This counter pauses when the channel becomes busy and resumes once idle again for a DIFS period. When the counter reaches zero, transmission may begin depending on the idleness of secondary channels over a shorter period (PIFS): under static channel bonding~(SCB), transmission is deferred if any secondary channel is busy, while dynamic channel bonding~(DCB) allows transmission on the primary and any contiguous idle secondary channels according to IEEE~802.11 bonding rules. After each unsuccessful transmission, its binary exponential backoff mechanism doubles the contention window~(CW). On a transmission success, the CW is reset to the minimum value.

\subsection{Multi-Armed Bandits (MABs)}
A MAB problem is defined by the tuple $\langle \mathcal{A}, \mathcal{R} \rangle$~\cite{rao_bandits}:
\begin{itemize}
    \item $\mathcal{A}$ is the \emph{action space}, a finite set of $k$ discrete arms.
    \item $\mathcal{R} = \{\mathcal{R}^a\}_{a \in \mathcal{A}}$ is a set of unknown reward distributions.
\end{itemize}

A CMAB problem incorporates side information (or \emph{context}) and is defined by the tuple $\langle \mathcal{A}, \mathcal{X}, \mathcal{R} \rangle$~\cite{rao_bandits}, where:
\begin{itemize}
    \item $\mathcal{A}$ is the \emph{action space}, a finite set of $k$ discrete arms.
    \item $\mathcal{X} \subseteq \mathbb{R}^d$ is the \emph{context space}, from which a $d$-dimensional feature vector $\mathbf{x}_t$ is observed at each round $t$.
    \item $\mathcal{R} = \{\mathcal{R}^a_\mathbf{x}\}_{a \in \mathcal{A}, \mathbf{x} \in \mathcal{X}}$ is a set of unknown context-conditional reward distributions.
\end{itemize}

The objective of a (C)MAB agent is to maximize the expected cumulative reward over $T$ rounds by sequentially selecting an action $a_t \in \mathcal{A}$ (conditioned on $\mathbf{x}_t$ in CMABs), observing a reward $r_t$, and updating its action-value estimates accordingly.

\subsection{Learning-based Channel Access}
\begin{figure}[t]
    \centering
    \includegraphics[width=\linewidth]{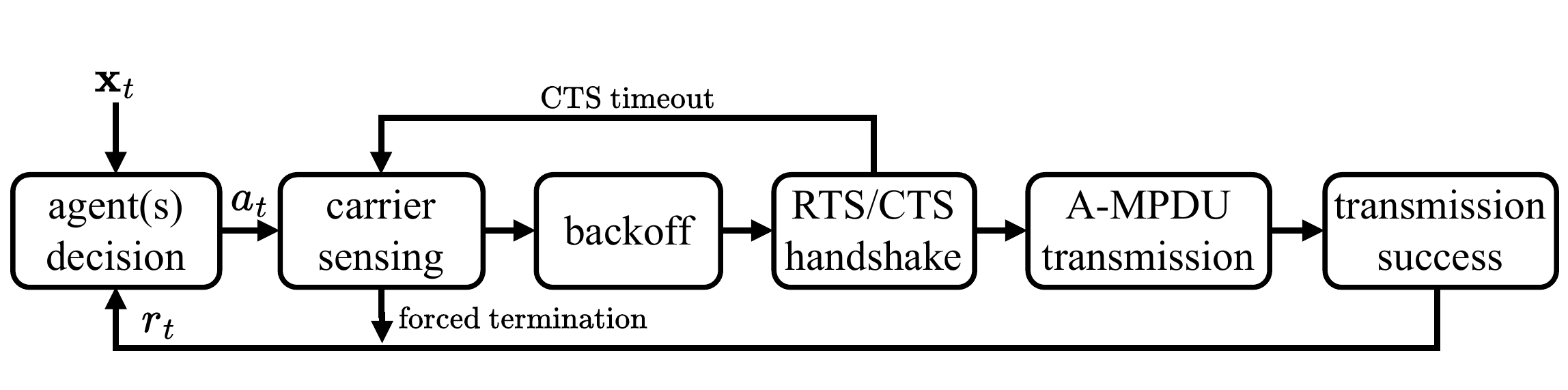}
    \caption{Learning-enabled access point transmission cycle.}
    \label{fig:tx_cycle}
\end{figure}

\begin{figure*}[t]
    \centering
    \includegraphics[width=0.95\linewidth]{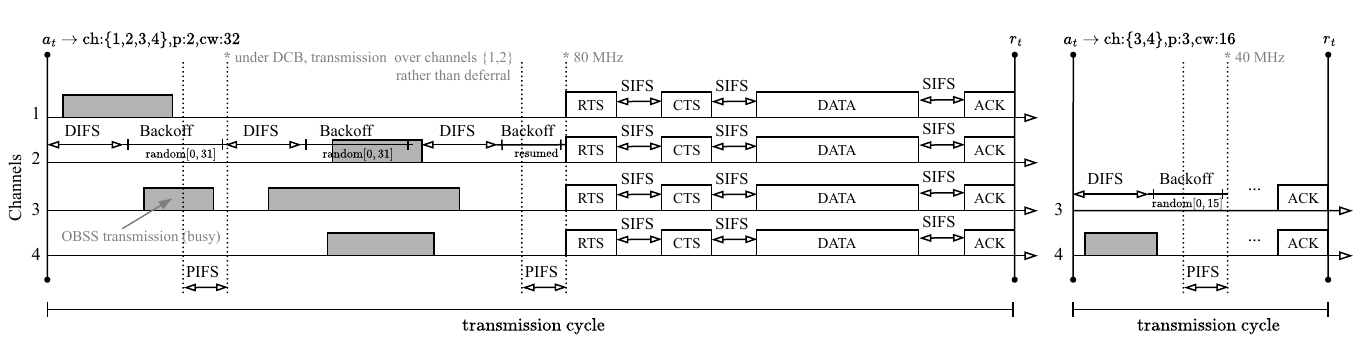}
    \caption{Temporal evolution of the channel access mechanism for a learning-enabled AP. The AP aims to access four channels (\{1, 2, 3, 4\}) using SCB. On the left, it contends multiple times on the primary channel (2); after two deferrals (one due to busy secondary channels at transmission time and one due to the primary channel becoming busy during backoff), it finally completes its backoff, finds all secondary channels idle during a PIFS period, and initiates transmission via an RTS/CTS handshake. On the right, the AP undergoes the same procedure but over channels \{3, 4\} (primary = 3).}
    \label{fig:tx_evo}
\end{figure*}

According to the framework introduced in~\cite{casasnovas2025paper1}, we model Wi-Fi channel access as a sequential learning game in which, at the beginning of each round $t$, each learning-enabled AP selects its transmission configuration, including the \emph{operational channel} (defined as the set of contiguous 20~MHz basic channels configured for transmission and thus determining the channel width), the \emph{primary channel}, and the \emph{contention window~(CW)}. Each learning round corresponds to a \emph{transmission cycle}, beginning with carrier sensing and ending upon successful acknowledgment or timeout, as illustrated in Fig.~\ref{fig:tx_cycle} and Fig.~\ref{fig:tx_evo}. To avoid persisting with inefficient configurations, a cycle is forcefully terminated if the AP cannot obtain a transmission opportunity for longer than $D_{\max}=10$~ms (approximately the 80th percentile latency of a Wi-Fi hop~\cite{sui2016characterizing}). No binary exponential backoff is applied, as the CW is learned online.


\medskip
\textbf{Actions ($\mathcal{A}$):}  
In each round, the learning AP selects an action vector $a_t = (a^{\text{ch}}, a^{\text{p}}, a^{\text{cw}})$:
\begin{itemize}
    \item $a^{\text{ch}} \in \{ \{1\}, \{2\}, \{3\}, \{4\}, \{1,2\}, \{3,4\}, \{1,2,3,4\} \}$ is the operational channel and determines the channel width (20/40/80~MHz);\footnote{A total of four 20-MHz basic channels are considered, identified as $\{1,2,3,4\}$. Receivers monitor all when idle, allowing transmitters to switch on a transmission basis.}
    \item $a^{\text{p}} \subseteq a^{\text{ch}}$ is the primary channel selected from the configured operational channel and where the backoff will be run;
    \item $a^{\text{cw}} \in \{2^{i+4} \mid i=0,\dots,6\}$ is the CW size.
\end{itemize}

Actions $a\in \mathcal{A}$ (where $K = |\mathcal{A}|$ is the total number of actions) are selected either (i) \emph{jointly} by a single agent~(SA), yielding $K^{\text{SA}} = 84$, 
or (ii) \emph{independently} by multiple agents~(MA), each responsible for one parameter and operating sequentially: operational channel agent ($K^{\text{ch}} = 7$), primary channel agent ($K^{\text{p}} \leq 4$), and CW agent ($K^{\text{cw}} = 7$). 

\medskip
\textbf{Reward ($\mathcal{R}$):} At the end of each learning round, a scalar reward $r\in[0,1]$ is computed based on the transmission cycle duration $D$ using min--max normalization and clipping:
\begin{equation*}
r = \text{clip}_{[0,1]}\!\left(\frac{D_{\max} - D}{D_{\max} - D_{\min}}\right),
\end{equation*}
with $D_{\min}=0$~ms and $D_{\max}=10$~ms, encouraging faster and more efficient channel access. In MA settings, all agents are given the same reward value, encouraging cooperative behavior.

\medskip
\textbf{Context ($\mathcal{X}$):} In CMABs, each agent observes a normalized feature vector $\mathbf{x}\!\in\![0,1]^d$, including: occupancy ratios per 20~MHz channel~$[0,1]^4$ (averaged over a 100~ms sliding window); instantaneous busy/idle flags per 20~MHz channel~$\{0,1\}^4$; and MAC-layer transmission queue utilization~$[0,1]$ (excluded for the primary agent). In MA setups, later agents incorporate prior agents’ decisions through one-hot encodings~$\{0,1\}^4$, enabling implicit coordination: the primary agent receives the selected operational channel, while the CW agent receives both the selected operational and primary channels.


\section{(C)MAB Algorithms Under Study}
\label{sec:algorithms}

In this paper we evaluate four representative action-selection strategies that balance \emph{exploration} (trying uncertain actions to improve reward estimates) and \emph{exploitation} (choosing those with high estimated reward):  
a) \textit{UCB}, a non-contextual, optimism-based MAB;  
b) \textit{OSUB}, a non-contextual approach exploiting unimodal reward structures and building upon optimism-based principles;  
c) \textit{LinUCB}, a contextual, optimism-based bandit; and  
d) \textit{E-RLB} (proposed), a contextual bandit using randomized ($\epsilon$-greedy) exploration.

\subsection{Upper Confidence Bound (UCB)}

UCB~\cite{auer2002finite} uses confidence bounds that shrink as arms are sampled more frequently:
\begin{equation*}
    a_t = \arg\max_{a \in \mathcal{A}} \left[ \hat{\mu}_a(t) + \sqrt{\frac{\alpha \ln t}{2 N_a(t)}} \right],
\end{equation*}
where $\hat{\mu}_a(t)$ is the empirical mean reward of action $a$, $N_a(t)$ its selection count, and $\alpha>0$ controls exploration (e.g., $\alpha=4$ in UCB1). The computational complexity of UCB over a horizon $T$ is $\mathcal{O}(K T)$.


\subsection{Optimal Sampling for Unimodal Bandits (OSUB)}

OSUB~\cite{combes2014unimodal} assumes a \emph{unimodal} reward structure on a connected graph $G = (V, E)$, where vertices $V$ represent actions and edges $E$ define neighborhood relations.
This means that, for every action $a \in V$, there exists a path in $G$ to the optimal action along which the expected reward increases monotonically.  
In each round, OSUB identifies the current \emph{leader} as $L(t) = \arg\max_{a} \hat{\mu}_a(t)$ and restricts exploration to it and its neighbors. 
Each arm in this set is assigned a KL-UCB~\cite{garivier2011kl} index, and the arm with the highest index is selected, except when the number of times $L(t)$ has been the leader is a multiple of $(\gamma{+}1)$ times ($\gamma$ denotes the maximum node degree in $G$), in which case the leader itself is reselected.
This structure-aware sampling achieves $\mathcal{O}(K \log T)$ computational complexity.

\textbf{Implementation note:}
In our MA~OSUB implementation, we introduce a small random exploration factor ($p=0.05$). This factor is crucial to prevent agents operating under shared rewards from falling into identical update trajectories and repeatedly selecting the same joint action---a synchronization issue commonly observed in UCB-based multi-agent systems (see Section~\ref{sec:evaluation}). For MA~OSUB, the neighborhood graph is designed to reflect the inherent structure of each decision dimension:  
for primary channel and CW selection, we adopt a linear topology where each action is connected to its adjacent values, capturing their ordinal relationship;  
for operational channel assignment, neighbors are defined as those configurations sharing at least one 20~MHz basic channel, reflecting spectral overlap.  
Similarly, in the SA~OSUB case, neighbors are defined as joint actions that share a channel and have adjacent or equal primary and CW values, ensuring structural consistency across action dimensions.


\subsection{Linear UCB (LinUCB)}

LinUCB~\cite{li2010contextual} (in its disjoint variant) is a contextual adaptation of the classical UCB algorithm that assumes a disjoint linear reward model $\mathbb{E}[r_t|\mathbf{x}_t]=\mathbf{x}_t^\top \boldsymbol{\theta}_a^*$ and selects:
\begin{equation*}
    a_t = \arg\max_{a \in \mathcal{A}} \!\left[ \hat{\boldsymbol{\theta}}_a^\top \mathbf{x}_t + \alpha \sqrt{\mathbf{x}_t^\top \mathbf{A}_a^{-1} \mathbf{x}_t} \right],
\end{equation*}
where $\hat{\boldsymbol{\theta}}_a$ is a ridge-regression estimate of $\boldsymbol{\theta}_a^*$ and $\mathbf{A}_a=\mathbf{D}_a^\top \mathbf{D}_a+\mathbf{I}_d$, with $\mathbf{D}_a \in \mathbb{R}^{t \times d}$ denoting the design matrix---whose $i$-th row is the context vector observed when action $a$ was selected at time $i$---and $\mathbf{I}_d$ being the $d \times d$ identity matrix.  
Its computational complexity is $\mathcal{O}(K d^2 T)$, where $d$ is the context dimensionality.

\subsection{Epsilon-RMSProp Linear Bandit (E-RLB)}

\SetKwInput{KwInit}{Initialize}
\begin{algorithm}[t]
\footnotesize
\setstretch{1.3}
\caption{E-RLB}
\label{alg:eps_rmsprop}
\KwIn{$\epsilon \in [0,1]$, $\eta > 0$, $\gamma \in (0,1]$, $\alpha_{\mathrm{ema}} \in [0,1)$, $\varepsilon>0$}
\KwInit{$\hat{\boldsymbol{\theta}}_a \gets \mathbf{0}_d$, $\hat{\boldsymbol{\theta}}_a^{\mathrm{ema}} \gets \mathbf{0}_d$, $v_a \gets \mathbf{0}_d$;  $\forall a \in \mathcal{A}$}

\For{$t = 1, 2, \dots$}{
    Observe context $\mathbf{x}_t \in \mathbb{R}^d$\;
    
    Select $a_t \gets 
        \begin{cases}
        \text{random arm in } \mathcal{A}_t & \text{with prob. } \epsilon \\
        \arg\max\limits_{a \in \mathcal{A}_t} \mathbf{x}_t^\top \hat{\boldsymbol{\theta}}_a^{\mathrm{ema}} & \text{otherwise}
        \end{cases}$\;
    \vspace{0.4em}
    Observe reward $r_t$\;
    
    $g_t \gets (\mathbf{x}_t^\top \hat{\boldsymbol{\theta}}_{a_t} - r_t) \cdot \mathbf{x}_t$\;

    $v_{a_t} \gets \gamma \cdot v_{a_t} + (1 - \gamma) \cdot g_t^2$\;
    \vspace{0.4em}
    $\displaystyle\hat{\boldsymbol{\theta}}_{a_t} \gets \hat{\boldsymbol{\theta}}_{a_t} - {\eta}/{\sqrt{v_{a_t} + \varepsilon}} \cdot g_t$\;
    \vspace{0.4em}
    $\hat{\boldsymbol{\theta}}_{a_t}^{\mathrm{ema}} \gets \alpha_{\mathrm{ema}} \cdot \hat{\boldsymbol{\theta}}_{a_t}^{\mathrm{ema}} + (1 - \alpha_{\mathrm{ema}}) \cdot \hat{\boldsymbol{\theta}}_{a_t}$\;
}
\end{algorithm}

Our newly proposed E-RLB algorithm, summarized in Algorithm~\ref{alg:eps_rmsprop}, combines $\epsilon$-greedy exploration with RMSProp-based updates under a disjoint linear realizability assumption, where each arm’s expected reward is a linear function of the context.   
In each round, with probability $\epsilon$, a random action is selected (exploration); otherwise, the arm with the highest estimated reward is chosen (exploitation):
\begin{equation*}
    a_t = \arg\max_{a \in \mathcal{A}} \mathbf{x}_t^\top \hat{\boldsymbol{\theta}}_a^{\mathrm{ema}},
\end{equation*}
where $\hat{\boldsymbol{\theta}}_a^{\mathrm{ema}}$ is an exponential moving average~(EMA) of the parameter estimates:
\begin{equation*}
    \hat{\boldsymbol{\theta}}_a^{\mathrm{ema}} \leftarrow \alpha_{\mathrm{ema}} \cdot \hat{\boldsymbol{\theta}}_a^{\mathrm{ema}} + (1 - \alpha_{\mathrm{ema}}) \cdot \hat{\boldsymbol{\theta}}_a.
\end{equation*}
The raw parameters $\hat{\boldsymbol{\theta}}_a$ are updated online via RMSProp~\cite{hinton2012lec6}:
\begin{equation*}
    g_{t} = (\mathbf{x}_t^\top \hat{\boldsymbol{\theta}}_a - r_t) \cdot \mathbf{x}_t,
\end{equation*}
\begin{equation*}
    v_{a,t} = \gamma \cdot v_{a,t{-}1} + (1 - \gamma) \cdot g_{t}^2,
\end{equation*}
\begin{equation*}
    \hat{\boldsymbol{\theta}}_a \leftarrow \hat{\boldsymbol{\theta}}_a - \frac{\eta}{\sqrt{v_{a,t} + \varepsilon}} \cdot g_{t},
\end{equation*}
with learning rate $\eta$, decay $\gamma$, and small $\varepsilon>0$ for numerical stability. Here, $g_t$ denotes the prediction error gradient and $v_{a,t}$ the RMSProp accumulator. Its computational complexity is $\mathcal{O}(K d T)$.

\textbf{Motivation note:} E-RLB is a lightweight, contextual linear bandit designed for non-stationary environments and relying on randomized exploration.
Unlike contextual optimism-based methods such as LinUCB, which are computationally demanding in high-dimensional contexts, E-RLB remains computationally efficient while preserving linear representational power. 
Its EMA estimates of arm-specific reward parameters implement exponential forgetting of past observations, allowing the algorithm to adapt to recent changes in the environment, a crucial feature to address the non-stationarity of wireless networks.
Its stochastic gradient updates (SGD) with RMSProp\footnote{RMSProp divides the learning rate by an exponentially decaying average of squared gradients, enhancing stability and accelerating convergence. Other popular optimizers include Adagrad, Adadelta, and Adam; RMSProp resolves Adagrad's diminishing learning rate issue~\cite{hinton2012lec6}.} minimize the squared prediction error of these parameters while providing adaptive per-parameter learning rates, improving convergence and stability.
While E-RLB is a heuristic conceived to perform well in practical, non-stationary deployments and no formal theoretical guarantees can be provided, its empirical performance is validated through simulations in Section~\ref{sec:evaluation}. Nevertheless, it is grounded in the standard contextual linear bandit framework, which achieves sublinear regret $\tilde{\mathcal{O}}(\sqrt{K d T})$ in stationary settings~\cite{li2010contextual}, and uses $\epsilon$-greedy exploration, whose worst-case regret is linear, $\mathcal{O}(T)$, in stationary settings.


\section{Performance Evaluation}
\label{sec:evaluation}

\begin{figure}[t]
    \centering
    \includegraphics[width=.8\linewidth]{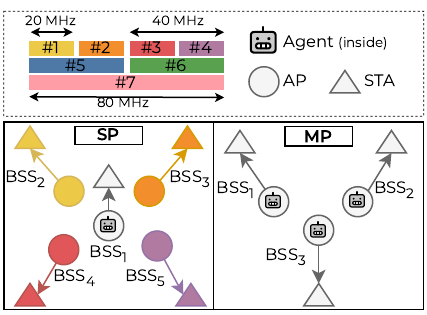}
    \caption{Single-Player (SP) and Multi-Player (MP) scenarios.}
    \label{fig:scenarios}
\end{figure}

We evaluate the (C)MAB algorithms from Section~\ref{sec:algorithms} using \texttt{WiPySim}\footnote{\url{https://github.com/miguelcUPF/WiPySim}} in the scenarios illustrated in Fig.~\ref{fig:scenarios} (see Table~\ref{tab:simparams} for simulation parameters and Table~\ref{tab:hyperparams} for tuned hyperparameters, tuned following the procedure described in \cite{casasnovas2025paper1}). Operational channels are labeled \#1--\#7, as illustrated in Fig.~\ref{fig:scenarios}, and the primary channel (when relevant) is indicated as a subscript (e.g., \#7$_3$ denotes the 80~MHz allocation \{1, 2, 3, 4\} with primary channel~3). 

All algorithms are assessed under SA and MA architectures, with agents in MA using the same algorithm and hyperparameters. Results for baseline configurations, representing IEEE~802.11 legacy operation, are available in~\cite{casasnovas2025paper1}.

Simulations run for 60~seconds and are repeated over 20~independent trials for statistical validation.
The considered scenarios involve multiple BSSs competing for the medium in a \emph{fully decentralized and non-cooperative} manner, each including a single transmitter--receiver pair with downlink traffic only, fixed positions, modulation and coding scheme~(MCS)~11 (i.e., 1024-QAM with 5/6 coding rate), and mutual coverage. All BSSs operate under a common channel-bonding mode, and each scenario is evaluated under both SCB and DCB. In particular, the considered single- and multi-player scenarios correspond to SP$2$ and MP$1$ in~\cite{casasnovas2025paper1}, serving as representative examples of the performance of all algorithm--architecture pairs. The simulation outputs for all the studied scenarios, including those shown in~\cite{casasnovas2025paper1}, are available in the associated dataset~\cite{dataset_miguel}.

Performance is evaluated in terms of: goodput $\Gamma_i$, defined as the rate of successfully delivered application-layer data in BSS$_i$; delay $d_i$, the per-packet latency including queuing, backoff, and (re)transmission times (up to 7 retransmissions); and fairness $\mathcal{J}$, measured using Jain's index~\cite{jain1984quantitative} over a BSS set $\mathcal{S}$, i.e., $\mathcal{J}_{\mathcal{S}} = (\sum_i \overline{\Gamma}_i)^2 / (|\mathcal{S}| \sum_i \overline{\Gamma}_i^2)$, where $\overline{(\cdot)}$ denotes the mean operator.

\begin{table}[t]
\centering
\caption{Simulation parameter settings.}
\setlength{\tabcolsep}{4pt}
\footnotesize
\label{tab:simparams}
\begin{tabular}{@{}llll@{}}
\toprule
\textbf{Parameter} & \textbf{Value} & \textbf{Parameter} & \textbf{Value} \\
\midrule
Carrier frequency & 5 GHz & PHY/MAC & IEEE 802.11ax\\
No. basic channels & 4  & Bandwidth & 20/40/80 MHz \\
MCS index & 11 & Spatial streams & 2 \\
Tx power & 20 dBm & Gain (Tx/Rx) & 0 dB / 0 dB \\
RTS/CTS & Enabled & Tx queue size & 500 packets \\
max A-MPDU size & 65,535 B & CW (min/max) & 16 / 1024 \\
Retry limit & 7 & Channel bonding & Static/Dynamic \\
Path loss model & See~\cite{casasnovas2025paper1} & Packet error rate & 0.1 \\
\bottomrule
\end{tabular}
\end{table}

\begin{table}[t]
    \centering
    \caption{Tuned hyperparameters.}
    \label{tab:hyperparams}
    \footnotesize
    \begin{tabular}{@{}l c c c c c c c c@{}}
    \toprule
     & \multicolumn{1}{c}{UCB
     } 
     & \multicolumn{1}{c}{LinUCB
     } & \multicolumn{4}{c}{E-RLB
     } 
     \\
    \cmidrule(lr){2-2} \cmidrule(lr){3-3} \cmidrule(lr){4-7}
     & $\alpha$ 
     & $\alpha$ & $\epsilon$ & $\eta$ & $\gamma$ & $\alpha_{\text{ema}}$ \\
    \midrule
    SA & 1.09 
    & 0.52  & 0.020 & 0.086 & 0.87 & 0.22 \\
    MA & 1.14 
    & 0.50  & 0.038 & 0.069 & 0.79 & 0.25 \\
    \bottomrule
    \end{tabular}
    \vspace{1mm}
    \begin{tablenotes}
    \footnotesize
    \item $\alpha$ (UCB) $ \in (1.0, 10.0)$; $\alpha$ (LinUCB) $\in (0.2, 20.0)$; $\epsilon \in (0.01, 0.30)$, $\eta \in (10^{\text{-}4}, 10^{\text{-}1})$ log scale, $\gamma \in (0.70, 0.99)$, $\alpha_{\text{ema}} \in (0.01, 0.30)$
    \end{tablenotes}
\end{table}

\subsection{Single-Player Performance}

The single-player scenario evaluates the algorithms in a setting where rewards are independent of other players’ actions. A single learning-enabled BSS (BSS$_1$) coexists with four legacy IEEE 802.11 BSSs (BSS$_2$--BSS$_5$), each on a distinct 20~MHz channel. The learning AP has full-buffer traffic, while legacy APs generate dynamic, non-saturated traffic varying every 15~seconds. To evaluate the learning strategies' ability to effectively adapt to changing conditions, in each interval one legacy AP's load drops to 10--20\%, providing a transient opportunity for BSS$_1$, while the others remain near 80--90\%. 

\medskip
\noindent\subsubsection{Static Channel Bonding (SCB)}$\;$
\label{sec:singleplayer_scb}

\begin{figure}[]
    \centering
    \begin{subfigure}[t]{0.95\linewidth}
        \centering
        \includegraphics[width=\linewidth]{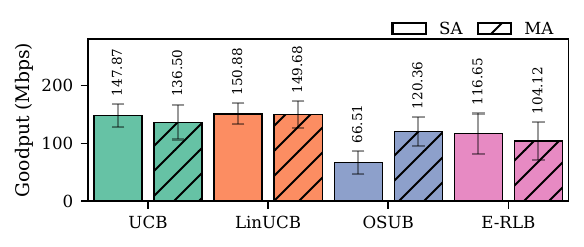}
        \caption{Aggregate across all trials.}
        \label{fig:scenarioB_global}
    \end{subfigure}
    \begin{subfigure}[t]{0.95\linewidth}
        \centering
        \includegraphics[width=\linewidth]{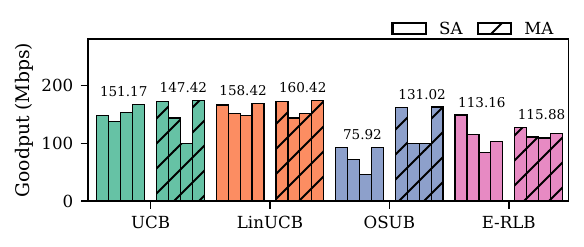}
        \caption{Representative trial (sub-bars correspond to 15-second intervals; displayed values represent the average goodput across all intervals).}
        \label{fig:scenarioB_trial}
    \end{subfigure}
    \caption{BSS$_1$ goodput under SCB in the single-player scenario.}
    \label{fig:scenarioB_combined}
\end{figure}

\begin{figure}[]
    \centering
    \begin{subfigure}[t]{0.95\linewidth}
        \centering
        \includegraphics[width=\linewidth]{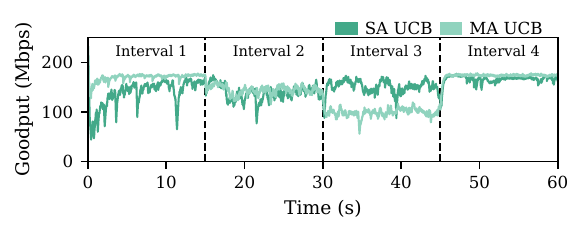}
        \caption{UCB.}
        \label{fig:B_ucb}
    \end{subfigure}
    \begin{subfigure}[t]{0.95\linewidth}
        \centering
        \includegraphics[width=\linewidth]{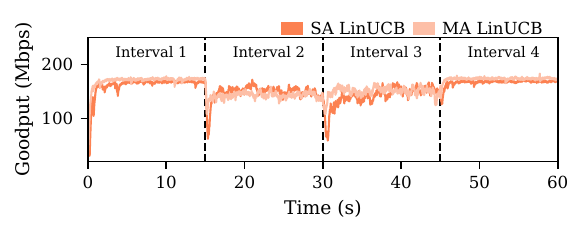}
        \caption{LinUCB.}
        \label{fig:B_linucb}
    \end{subfigure}
    \begin{subfigure}[t]{0.95\linewidth}
        \centering
        \includegraphics[width=\linewidth]{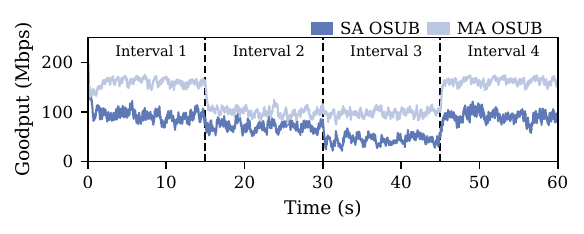}
        \caption{OSUB.}
        \label{fig:B_osub}
    \end{subfigure}
    \begin{subfigure}[t]{0.95\linewidth}
        \centering
        \includegraphics[width=\linewidth]{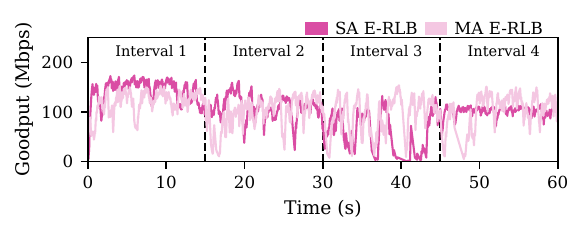}
        \caption{E-RLB.}
        \label{fig:B_erlb}
    \end{subfigure}
    \caption{Goodput evolution of BSS$_1$ under SCB in the single-player scenario for a representative trial.}
    \label{fig:scenarioB}
\end{figure}

Figures ~\ref{fig:scenarioB_global} and~\ref{fig:scenarioB_trial} report the average goodput (and standard deviation) achieved by BSS$_1$ under SCB for each algorithm--architecture pair, across all trials and for a representative trial, respectively. Fig.~\ref{fig:scenarioB} further illustrates the temporal evolution of BSS$_1$'s goodput in the representative trial (but the observed trends are consistent across all runs). In this trial, the optimal channel selections for BSS$_1$ are \#1, \#2, \#3, and \#1, respectively, across intervals. Packet delay is omitted from the figures, as its behavior closely mirrors the goodput results---a correlation also highlighted in~\cite{casasnovas2025paper1}.

The performance of the algorithms is summarized next.

\textbf{UCB} achieves robust performance across trials (SA: $\overline{\Gamma}_1\!=\!147.9$~Mbps, $\overline{d}_1\!=\!35.5$~ms; MA: $\overline{\Gamma}_1\!=\!136.5$~Mbps, $\overline{d}_1\!=\!40.0$~ms). In particular, as observed in the representative trial, SA~UCB attains high goodput across the four intervals ($\overline{\Gamma}_1\!=\!147.7$~Mbps, $137.4$~Mbps, $152.8$~Mbps, and $166.8$~Mbps, respectively), as it consistently selects the interval-wise optimal channels (\#1: 86.4\% in the 1st, \#2: 86.5\% in the 2nd, \#3: 96.7\% in the 3rd, \#1: 98.5\% in the 4th). In contrast, MA~UCB, converges faster initially but performs poorly in the 3rd interval ($\overline{\Gamma}_1\!=\!171.8$~Mbps in the 1st, $143.8$~Mbps in the 2nd, $100.0$~Mbps in the 3rd, and $174.1$~Mbps in the 4th). In particular, although it selects the optimal allocations in most intervals (\#1: 99.4\% in the 1st, \#2: 79.3\% in the 2nd, \#1: 100\% in the 4th), it remains on channel \#1 in the 3rd instead of switching to \#3.
This behavior arises from the combination of the standard UCB initialization (each action is initially pulled once) and the shared reward structure in the MA setting. Since agents share a common, non-separable reward, their empirical estimates evolve identically, tightly coupling their actions and hindering independent exploration. In our MA formulation, actions are initialized sequentially within each agent's action space; since the channel and CW action sets have the same cardinality, this implicitly pairs channels and CWs with matching indices. As a result, lower-indexed channels are coupled with smaller CWs and higher-indexed channels with larger CWs, introducing an inherent bias toward actions involving lower-indexed channels. This bias leads to performance degradation whenever channels~\#3 or~\#4 constitute the interval-wise optimum, as the agent may favor a lower-indexed channel (e.g., \#1) due to shorter waiting times despite higher contention. Notice that randomized initialization would only produce alternative pairings and would not resolve the inherent coupling problem.

\textbf{LinUCB} achieves the best overall performance (SA: $\overline{\Gamma}_1\!=\!150.9$~Mbps, $\overline{d}_1=34.7$~ms; MA: $\overline{\Gamma}_1\!=\!149.7$~Mbps, $\overline{d}_1=35.4$~ms). In the representative trial, both SA and MA variants sustain high goodput across intervals ($\overline{\Gamma}_1\!=\!166.3/172.6$~Mbps, $151.1/143.6$~Mbps, $147.7/151.6$~Mbps, and $168.5/173.4$~Mbps, respectively), since they
reliably select the optimal channel in each interval (\#1: 99.7\%/100\% in the 1st, \#2: 99.7\%/100\% in the 2nd, \#3: 97.4\%/99.9\% in the 3rd, \#1: 99.9\%/100\% in the 4th, SA/MA). In particular, the MA variant converges slightly faster than the SA variant.

Importantly, compared to UCB, LinUCB converges faster and more reliably, selecting the optimal channel in more than 96\%/99\% of decisions (SA/MA) across trials in the 4th interval, as contextual information enables rapid re-identification of optimal actions previously observed under similar conditions. Note that, in the representative trial, both SA and MA~UCB also converge quickly and frequently select the optimal channel in the 4th interval (the latter mainly because it remained on channel \#1 in the previous interval). However, in other trials, SA~UCB selects the 4th-interval optimum in only about $70\%$ of decisions, whereas MA~UCB often fails to identify it. Moreover, MA~LinUCB avoids the joint-action coupling observed in MA~UCB, as each agent conditions its decisions on independent contextual inputs, thereby maintaining exploration diversity.

\textbf{OSUB} attains moderate performance in the MA setting but performs significantly worse in the SA setting (SA: $\overline{\Gamma}_1\!=\!66.5$~Mbps, $\overline{d}_1\!=\!87.7$~ms; MA: $\overline{\Gamma}_1\!=\!120.4$~Mbps, $\overline{d}_1\!=\!44.3$~ms). 
In particular, as observed in the representative trial, SA~OSUB achieves degraded goodput across intervals, especially in the second and third interval ($\overline{\Gamma}_1\!=\!92.8$~Mbps in the 1st, $72.2$~Mbps in the 2nd, $46.3$~Mbps in the 3rd, and $92.4$~Mbps in the 4th). In contrast, MA~OSUB converges optimally and fast in the 1st and 4th intervals, as it frequently selects the optimal channel (\#1: 96.6\% in both cases), but underperforms in the 2nd and 3rd intervals ($\overline{\Gamma}_1\!=\!161.7$~Mbps in the 1st, $100.2$~Mbps in the 2nd, $99.4$~Mbps in the 3rd, and $162.8$~Mbps in the 4th).
The reason both variants exhibit pronounced degradation in the 2nd and 3rd intervals is due to OSUB's unimodal leader-neighbor structure and the adopted channel-neighboring rule (neighbors must share at least one 20~MHz channel). In particular, in the trial considered, channel \#1 consistently becomes the leader and its adjacent neighbors (\#5, \#7) underperform under SCB (as observed in~\cite{casasnovas2025paper1}). This prevents exploration of non-adjacent channels (e.g., \#2, \#3), thereby violating the unimodality assumption on which OSUB relies. Consequently, performance degrades whenever the optimal channel differs from that of the 1st interval (note that if the optimal channel remained constant, MA~OSUB would achieve near-optimal performance).

On the other hand, SA~OSUB's larger degradation arises from its higher-dimensional joint action graph, which yields sparser sampling (fewer observations per joint action) and a higher node degree (more neighbors per node). Combined with the high variance of observed rewards, this increases uncertainty and causes OSUB's KL-UCB optimism mechanism to overexplore high-uncertainty combinations instead of consistently exploiting the true optimum (e.g., in the 1st interval, SA~OSUB selects channel \#1 only 50.99\% of the time, and \#5 and \#7 in 33.0\% and 15.6\%, respectively). In contrast, the channel agent in MA~OSUB operates over a smaller, smoother graph with fewer neighbors per node, favoring more stable learning.

The coupling observed in MA~UCB would also affect vanilla MA~OSUB; however, our MA~OSUB implementation includes a small random exploration factor, as described in Section~\ref{sec:algorithms}, allowing independent CW selection regardless of the chosen channel. This modification mitigates the coupling problem and enables MA~OSUB to outperform UCB in the 1st interval when the optimal channel is higher-indexed (e.g., \#3 or \#4), albeit at the cost of occasional transient performance drops due to (disruptive) neighboring-action exploration. 

\textbf{E-RLB} performs worse than other learning-based methods in all trials (SA: $\overline{\Gamma}_1\!=\!116.7$~Mbps, $\overline{d}_1=65.3$~ms; MA: $\overline{\Gamma}_1\!=\!104.12$~Mbps, $\overline{d}_1=68.0$~ms). In particular, as observed in the representative trial, both variants exhibit significant variability in goodput and frequent performance drops, leading to reduced interval-level performance ($\overline{\Gamma}_1\!=\!149.1/126.9$~Mbps in the 1st, $115.9/110.9$~Mbps in the 2nd, $99.4/109.1$~Mbps in the 3rd, and $126.9/116.7$~Mbps in the 4th, SA/MA). Indeed, although E-RLB tends to use the interval-wise optimal channels (\#1: 93.1\%/83.9\% in the 1st, \#2: 64.1\%/67.9\% in the 2nd, \#3: 32.1\%/53.8\% in the 3rd, \#1: 83.4\%/62.7\% in the 4th, SA/MA), its inherent uniform random exploration occasionally drives selections toward overloaded channels or excessively large CWs, leading to sharp transient performance drops. This behavior is particularly evident in the 2nd and 3rd intervals, where the combined effects of random exploration and parameter readjustments hinder consistent selection of the interval-wise optimum. 

In contrast, in the 4th interval, both SA and MA variants achieve higher stability and more frequent optimal selections, as the underlying conditions have already been observed and effectively incorporated into the learned parameters.
Indeed, both variants demonstrate effective learning by the end of the simulation; for instance, the learned EMA-based coefficients for channel-occupancy features are strongly negative for each action's corresponding channel, meaning that higher observed occupancy on that channel directly reduces its predicted reward (e.g., in MA~E-RLB, for the action selecting channel~\#1, the independently learned coefficients $(-2.1, -0.2, 0.1, -0.1)$ for the occupancy ratios of channels~\#1--\#4, significantly penalize this action when channel~\#1 is highly occupied). 

Notably, MA~E-RLB performs worse than its SA counterpart because its tuned exploration rate is higher, leading to more frequent disruptive exploration events and reduced stability. Even with the same exploration rate, the probability that at least one of the $n$ agents explores an action dimension, $1 - (1-\epsilon)^n$, always exceeds the single-agent probability $\epsilon$, for all $n > 1$ and $\epsilon \in (0,1)$. Thus, randomized exploration has an amplified impact in multi-agent settings.

\medskip
\noindent\subsubsection{Dynamic Channel Bonding (DCB)}$\;$

\begin{figure}[t!]
    \centering
    \begin{subfigure}[t]{0.95\linewidth}
        \centering
        \includegraphics[width=\linewidth]{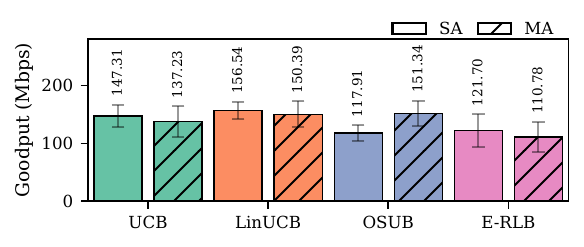}
        \caption{Aggregate across all trials.}
        \label{fig:scenarioB_dcb}
    \end{subfigure}
    \begin{subfigure}[t]{0.95\linewidth}
        \centering
        \includegraphics[width=\linewidth]{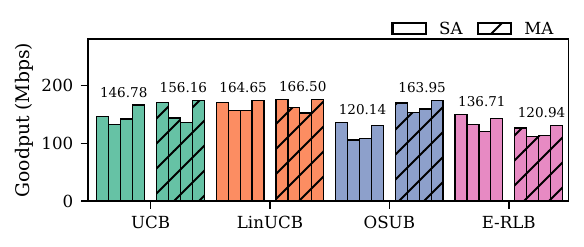}
        \caption{Representative trial (sub-bars correspond to 15-second intervals; displayed values represent the average goodput across all intervals).}
        \label{fig:scenarioB_dcb_trial}
    \end{subfigure}%
    \caption{BSS$_1$ goodput under DCB in the single-player scenario.}
    \label{fig:scenarioB_dcb_aggregate}
\end{figure}

Figures~\ref{fig:scenarioB_dcb} and~\ref{fig:scenarioB_dcb_trial} report the average goodput (and standard deviation) achieved by BSS$_1$ under DCB across all trials and for the representative trial considered in Section~\ref{sec:singleplayer_scb}, respectively.

Overall, all learning algorithms demonstrate effective adaptation under DCB, consistently prioritizing configurations using the underloaded channel as primary. Indeed, under DCB, primary channel selection becomes critical, as it exploits DCB's flexibility, enabling wider allocations to perform competitively. Consequently, under DCB, learning approaches achieve comparable or improved performance relative to SCB. 

\textbf{UCB} achieves similar overall performance to that under SCB (SA: $\overline{\Gamma}_1 = 147.3$~Mbps, $\overline{d}_1 = 35.5$~ms; MA: $\overline{\Gamma}_1 = 137.2$~Mbps, $\overline{d}_1 = 39.6$~ms). As observed in the representative trial, in SA~UCB, selections are distributed across multiple configurations using the underloaded channel as primary, reflecting a non-convex utility function with multiple local optima (e.g., 1st interval: \#1~30.6\%, \#5$_1$~29.5\%, \#7$_1$~33.8\%; 3rd interval: \#3~30.3\%, \#6$_3$~32.5\%, \#7$_3$~31.7\%). In contrast, MA~UCB consistently converges to the optimal primary-only 20~MHz allocations across intervals, yielding substantially higher goodput in the representative trial than under SCB. Specifically, in the 3rd interval of the representative trial, MA~UCB correctly converges into the optimal primary channel, selecting \#3 in 86.7\% of decisions, while under SCB it remained using \#1. This highlights the stronger guiding effect of primary-channel selection under DCB. Nevertheless, across all trials, MA~UCB performance remains comparable to SCB due to the coupling of higher-indexed 20~MHz channels with larger CW values, which limits the benefits of wider channel opportunities.

\textbf{LinUCB} again achieves a similar performance than in SCB (SA: $\overline{\Gamma}_1 = 156.4$~Mbps, $\overline{d}_1 = 33.1$~ms; MA: $\overline{\Gamma}_1 = 150.4$~Mbps, $\overline{d}_1 = 35.1$~ms). As observed in the representative trial, both SA and MA variants prioritize configurations using the interval-wise optimal primary (e.g., for SA~LinUCB, 1st interval: \#1~22.1\%, \#5$_1$~62.2\%, \#7$_1$~16.2\%; 3rd interval: \#3~35.3\%, \#5$_3$~63.4\%).

\textbf{OSUB} achieves substantially higher performance than under SCB (SA: $\overline{\Gamma}_1 = 117.8$~Mbps, $\overline{d}_1 = 44.2$~ms; MA: $\overline{\Gamma}_1 = 151.3$~Mbps, $\overline{d}_1 = 34.8$~ms). This improvement arises because 40~MHz and 80~MHz allocations now deliver robust goodput under optimal primary-channel assignments (as observed in ~\cite{casasnovas2025paper1}), enabling effective navigation of the channel-assignment graph and consistent selection of configurations using the optimal channel as primary, as observed in the representative trial (e.g., for SA~OSUB, 1st interval: \#1~21.8\%, \#5$_1$~33.9\%, \#7$_1$~34.5\%; 3rd interval: \#3~18.0\%, \#6$_3$~27.4\%, \#7$_3$~35.3\%). MA~OSUB again outperforms SA~OSUB, as it consistently selects the smallest CW value (over 97\% of decisions per interval), benefiting from earlier channel access. In contrast, SA~OSUB explores multiple CW sizes (up to 256) due to increased node degrees, occasionally degrading its performance.

\textbf{E-RLB} also improves under DCB (SA: $\overline{\Gamma}_1 = 121.7$~Mbps, $\overline{d}_1 = 46.1$~ms; MA: $\overline{\Gamma}_1 = 110.8$~Mbps, $\overline{d}_1 = 50.7$~ms). As in SCB, it learns effectively, primarily selecting configurations using the interval-wise optimal primary, as observed in the representative trial (e.g., for SA~E-RLB, 1st interval: \#1 36.1\%, \#5$_1$ 37.9\%, \#7$_1$ 23.4\%; 3rd interval: \#3 34.5\%, \#6$_3$ 27.2\%, \#7$_3$ 24.5\%). However, frequent exploration of non-optimal configurations and large CWs continues to limit its stability and performance relative to other algorithms. 

Notably, both LinUCB and E-RLB MA variants tend to avoid the 80~MHz allocation more often potentially due to the larger masked action space of the primary-channel agent respect to other allocations ($K^{\text{p}}=4$ versus $K^{\text{p}}=1$ for 80~MHz and 20~MHz operational channels, respectively).

\subsection{Multi-Player Performance}

The multi-player scenario evaluates the algorithms when multiple self-interested learning-enabled BSSs (BSS$_1$--BSS$_3$) coexist and contend for spectrum resources.

\medskip
\noindent\subsubsection{Static Channel Bonding (SCB)}$\;$
\label{sec:multiplayer_scb}

\begin{figure}[t!]
    \centering
    \begin{subfigure}[t]{0.95\linewidth}
        \centering
        \includegraphics[width=\linewidth]{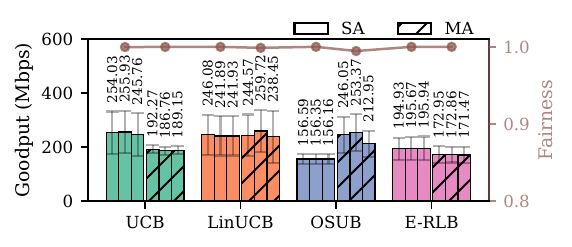}
        \caption{Aggregate across all trials.}
        \label{fig:scenarioC_global}
    \end{subfigure}
    \begin{subfigure}[t]{0.95\linewidth}
        \centering
        \includegraphics[width=\linewidth]{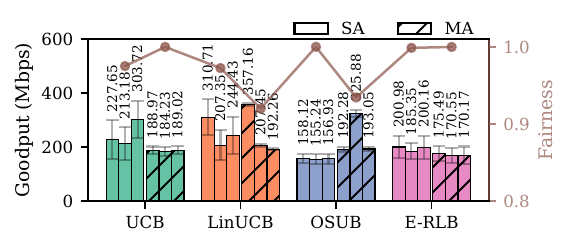}
        \caption{Representative trial.}
        \label{fig:scenarioC}
    \end{subfigure}%
    \caption{BSS goodput under SCB in the multi-player scenario. Brown markers show fairness across BSSs. Sub-bars correspond to BSS$_1$--BSS$_3$.}
    \label{fig:scenarioC_combined}
\end{figure}

Figures~\ref{fig:scenarioC_global} and~\ref{fig:scenarioC} report the goodput and fairness across BSSs under SCB, across all trials and for a representative trial, respectively. The observed trends in the representative trial are consistent with those across all runs.

Due to the non-convex shape of the utility function, the different algorithms lead the players to adopt several strategies, including: (i) two BSSs on the same 40~MHz channel with the remaining BSS on the other 40~MHz channel; (ii) each BSS on a distinct 20~MHz channel; (iii) one BSS on a 40~MHz channel and the other two on separate, contention-free 20~MHz channels; and (iv) one BSS on the 80~MHz channel, another BSS on a 40~MHz channel, and the remaining BSS on a 20~MHz channel (without overlapping the 40~MHz BSS).

\textbf{UCB} achieves the highest aggregate goodput across all trials in the SA setting, but low performance in MA ($\overline{\Gamma}_{1,2,3}\!=\!251.9/189.4$~Mbps\footnote{$\overline{\Gamma}_{1,2,3}$ denotes the average goodput across BSS$_1$--BSS$_3$.}, SA/MA). 
In particular, SA~UCB consistently converges to~(i), yielding high goodputs but reduced fairness, as the BSS operating alone on the 40~MHz channel achieves higher performance (e.g., in the representative trial: 
$\mathcal{J}\!=\!0.975$). 
In contrast, MA~UCB consistently converges to~(ii), likely due to joint-action coupling that biases agents toward lower-indexed 20~MHz allocations. This conservative strategy preserves fairness but underutilizes bandwidth, leading to decreased goodputs (e.g., in the representative trial: 
$\mathcal{J}\!\approx\!1$).

\textbf{LinUCB} achieves high average goodputs across all trials in both SA and MA settings ($\overline{\Gamma}_{1,2,3}\!=\!243.3/247.6$~Mbps, SA/MA).
In particular, SA~LinUCB consistently converges to~(i), offering high goodputs but reduced fairness (e.g., in the representative trial: 
$\mathcal{J}\!=\!0.973$). 
In contrast, MA~LinUCB consistently converges to~(iii), achieving high but uneven goodputs, as the BSS on the 40~MHz channel benefits from increased bandwidth (e.g., in the representative trial: 
$\mathcal{J}\!=\!0.920$).

\textbf{OSUB} achieves the lowest aggregate goodput across all trials in the SA setting, whereas MA~OSUB achieves high performance ($\overline{\Gamma}_{1,2,3}\!=\!156.4/237.4$~Mbps, SA/MA).  
In particular, SA~OSUB fails to reach a stable allocation due to high uncertainty, as observed in the single-player scenario: this uncertainty drives frequent exploration of channel configurations by individual BSSs, which in turn induces reactive adjustments by the others. Consequently, allocations continuously oscillate across patterns~(i)--(iv), significantly reducing overall goodput, while maintaining balanced per-BSS performance (e.g., in the representative trial, 
$\mathcal{J}\!\approx\!1$).
MA~OSUB, instead, mitigates these oscillations and consistently converges to a single allocation,~(iii), achieving high average goodput but degraded fairness, as the BSS on the 40~MHz channel benefits disproportionately (e.g., in the representative trial: 
$\mathcal{J}\!=\!0.934$).

\textbf{E-RLB} achieves intermediate aggregate goodputs across all trials ($\overline{\Gamma}_{1,2,3}\!=\!195.5/172.5$~Mbps, SA/MA).
Both SA and MA variants fail to reach a stable equilibrium, oscillating primarily between~(i) and~(iii) within each trial. Consequently, E-RLB achieves lower overall goodput than the other algorithms but maintains high fairness (e.g., in the representative trial, SA: 
$\mathcal{J}\!=\!0.999$).  
These oscillations arise from the algorithm's $\epsilon$-greedy exploration, which drives frequent reallocations and policy adjustments among BSSs: when one BSS switches channels, the others must adapt accordingly. In MA~E-RLB, this effect is amplified due to the higher aggregate exploration probability across agents, further reducing overall goodput.

\medskip
\noindent\subsubsection{Dynamic Channel Bonding (DCB)}$\;$

\begin{figure}[t!]
    \centering
    \begin{subfigure}[t]{0.95\linewidth}
        \centering
        \includegraphics[width=\linewidth]{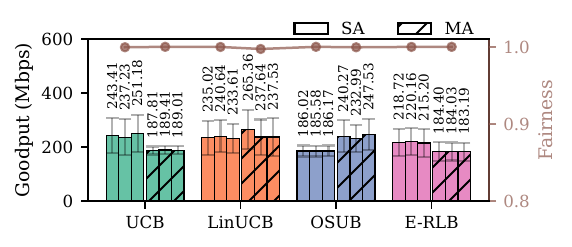}
        \caption{Aggregate across all trials.}
        \label{fig:scenarioC_dcb}
    \end{subfigure}
    \begin{subfigure}[t]{0.95\linewidth}
        \centering
        \includegraphics[width=\linewidth]{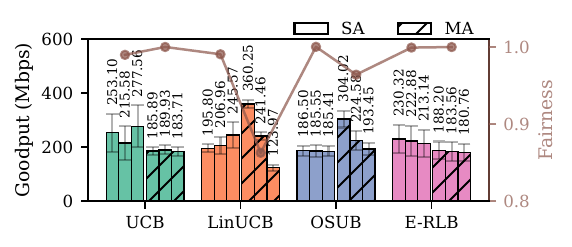}
        \caption{Representative trial.}
        \label{fig:scenarioC_dcb_trial}
    \end{subfigure}%
    \caption{BSS goodput under DCB in the multi-player scenario. Brown markers show fairness across BSSs. Sub-bars correspond to BSS$_1$--BSS$_3$.}
    \label{fig:scenarioC_dcb_combined}
\end{figure}

Figures~\ref{fig:scenarioC_dcb} and~\ref{fig:scenarioC_dcb_trial} report the goodput and fairness across BSSs under DCB, aggregated across all trials and for the representative trial considered in Section~\ref{sec:multiplayer_scb}, respectively.

Under DCB, each BSS may opportunistically exploit idle secondary channels beyond its primary one, increasing its transmission opportunities and effective bandwidth. This flexibility expands the number of favorable allocations compared to SCB:
(i) all BSSs on 80~MHz;
(ii) each BSS on a distinct 20~MHz channel;
(iii) two BSSs on 80~MHz and one on 40~MHz;
(iv) two BSSs on separate 40~MHz channels and one on 80~MHz;
(v) two BSSs on separate 40~MHz channels and one on 20~MHz; and
(vi) one BSS on 80~MHz, one on 40~MHz, and one on a 20~MHz channel (overlapping the 40~MHz BSS).  

Nevertheless, all algorithms generally assign distinct primary channels to BSSs, ensuring medium access (at least) via the primary whenever idle.

\textbf{UCB} again performs better in SA than in MA ($\overline{\Gamma}_{1,2,3}=243.9/188.7$~Mbps, SA/MA).  
SA~UCB typically converges to~(iii), occasionally to~(iv), thereby consistently utilizing high-bandwidth configurations for the BSSs and yielding more balanced goodput than under SCB (e.g., in the representative trial, where~(iv) is the most frequently selected configuration: $\mathcal{J}=0.990$).  
MA~UCB consistently converges to~(ii), mirroring its SCB behavior and thus achieving similar performance.

\textbf{LinUCB} achieves high performance in both architectures, closely matching SCB ($\overline{\Gamma}_{1,2,3}=236.4/246.8$~Mbps, SA/MA).  
SA~LinUCB primarily converges to~(iii), occasionally to~(i) (which may decrease performance as observed in the representative trial), again prioritizing high-bandwidth allocations.
Surprisingly, MA~LinUCB primarily alternates between~(v) and~(vi), achieving higher overall performance than SA~LinUCB despite using smaller-bandwidth configurations, potentially thanks to DCB's flexibility under overlapping allocations. In the representative trial, however, it places BSS$_2$ and BSS$_3$ on the same 40~MHz segment (with BSS$_3$ operating with a larger CW), while BSS$_1$ occupies the other 40~MHz segment; this isolated outcome significantly reduces fairness ($\mathcal{J}=0.863$).

\textbf{OSUB} improves clearly in SA and remains comparable in MA to its SCB results ($\overline{\Gamma}_{1,2,3}=185.9/240.3$~Mbps, SA/MA).  
SA~OSUB, as observed under SCB, fails to converge to a stable allocation, resulting in frequent oscillations among BSSs; however, high-bandwidth configurations are selected more often and it yields high fairness.
MA~OSUB primarily alternates between~(iii) and (iv), thereby continuing to use high-bandwidth configurations.

\textbf{E-RLB} maintains oscillatory behavior but achieves significantly higher goodput than under SCB ($\overline{\Gamma}_{1,2,3}=218.0/183.9$~Mbps, SA/MA).  
SA~E-RLB alternates mainly between~(iii) and~(iv).  
MA~E-RLB behaves similarly, primarily selecting~(iii) but exploring alternatives more frequently, reducing overall performance.

\section{Main Takeaways}\label{sec:takeaways}

Our results show that both the learning strategy and the action-space formulation strongly affect performance, convergence speed, and stability, confirming prior findings, such as the increased variability associated with randomized strategies and the implicit collaboration in multi-player settings due to shared interests (albeit with potentially reduced fairness). For instance, under DCB, learning-enabled BSSs consistently adopt non-overlapping primary channels, maximizing their effective transmission opportunities. 

In particular, our results indicate that decomposing the action space across multiple specialized agents~(MA) generally accelerates convergence relative to a single joint agent~(SA). However, under vanilla non-contextual methods, MA can exhibit inefficiencies: agents learning under shared rewards may follow identical update trajectories, repeatedly selecting the same joint action.
Optimism-driven methods (UCB, LinUCB) provide structured and stable exploration. LinUCB consistently achieves the highest steady-state performance and fastest re-convergence under recurrent conditions, highlighting the advantages of contextual learning. In contrast, vanilla UCB suffers in MA due to the joint-action coupling problem, a limitation that LinUCB avoids through independent, context-conditioned decisions.
Unimodal algorithms (OSUB) are highly sensitive to the correctness of the imposed unimodal structure. When the unimodality assumption does not hold, performance and stability degrade substantially. Simpler, locally connected graphs (as in MA~OSUB) improve stability and performance compared to higher-dimensional graphs (as in SA~OSUB), which increase uncertainty. Vanilla OSUB also suffers from joint-action coupling, though this can be mitigated through small randomized exploration at the cost of occasional disruptive reallocations.
Randomized exploration (as in E-RLB) via vanilla $\epsilon$-greedy induces frequent and sometimes disruptive joint actions. This significantly degrades performance, particularly in MA (due to higher aggregate exploration) and in multi-player settings (due to players mutually adapting to each other's exploratory actions). Nevertheless, E-RLB remains an effective learner, frequently converging to near-optimal configurations and avoiding persistent channel overlap. These results suggest that the lightweight, SGD-based approach of E-RLB is a viable low-complexity option, but its practical deployment requires enhancing its stability through more selective exploration strategies.

Overall, our findings emphasize that: 
\begin{itemize}
    \item LinUCB, a contextual, optimism-driven strategy, provides the most reliable performance and adaptation, particularly under recurrent conditions.  
    \item Unimodal methods require carefully designed graphs to satisfy the unimodality assumption; simpler, locally connected graphs improve stability.   
    \item Naive randomized exploration can destabilize learning and significantly degrade performance, especially in MA and multi-player settings.
    \item MA decompositions accelerate learning but increase sensitivity to randomized exploration and demand coordination under shared rewards; independent, context-aware decisions can mitigate MA correlated learning issues. 
    \item Our proposed contextual bandit, E-RLB, learns effectively but is negatively affected by randomized exploration.
\end{itemize}

\section{Conclusions} \label{sec:conclusions}

This paper evaluated MAB-based learning for Wi-Fi channel access, comparing joint (single-agent, SA) and factorial (multi-agent, MA) action-space formulations and contextual/non-contextual action-selection strategies: optimism-driven (UCB: state-of-the-art, non-contextual; LinUCB: state-of-the-art, contextual), unimodal (OSUB: state-of-the-art, non-contextual), and randomized (E-RLB: proposed, contextual).

Future work will focus on enhancing the stability and efficiency of E-RLB through selective exploration strategies, such as Thompson sampling or sticky $\epsilon$-greedy~\cite{carrascosa2020multi}. We will also investigate how increasing the dimensionality and complexity of the action space affects performance in SA versus MA formulations, and compare decentralized learning with centralized approaches in multi-player scenarios. Another promising direction is to assess whether stability improves when actions are held for multiple transmissions, guided by an adaptive meta-controller. Furthermore, contrasting MAB-based methods with stateful RL may highlight trade-offs between lightweight, online learning, and long-term policy adaptation. Finally, we plan to evaluate the proposed algorithms in more realistic environments, including mobility, interference dynamics, and denser deployments, where the probability of collisions is higher and CW adaptation may have a stronger impact. This assessment will be conducted through experimental testbeds and simulations to confirm robustness, analyze convergence properties, and establish practical applicability under real-world conditions.


\section{Acknowledgments}

This paper is supported by the CHIST-ERA Wireless AI 2022 call MLDR project (ANR-23-CHR4-0005), partially funded by AEI and NCN under projects PCI2023-145958-2 and 2023/05/Y/ST7/00004, respectively, by Wi-XR PID2021-123995NB-I00 and TRUE-Wi-Fi PID2024-155470NB-I00 (MCIU/AEI/FEDER,UE), by MCIN/AEI under the Maria de Maeztu Units of Excellence Programme (CEX2021-001195-M), and AGAUR ICREA Academia 00077.


\bibliographystyle{elsarticle-num} 
\bibliography{References}

\end{document}